\documentclass[superscriptaddress]{revtex4}%
\usepackage{amsfonts}
\usepackage{amsmath}
\usepackage{amssymb}
\usepackage{graphicx}
\providecommand{\U}[1]{\protect\rule{.1in}{.1in}}
\newtheorem{theorem}{Theorem}

\newtheorem{definition}[theorem]{Definition}

\begin{document}
\title{Predicting decoherence in discrete models}
\author{Mario Castagnino}
\affiliation{CONICET, IAFE (CONICET-UBA), IFIR and FCEN (UBA), Argentina}
\author{Sebastian Fortin}
\affiliation{CONICET, IAFE (CONICET-UBA) and FCEN (UBA), Argentina}

\begin{abstract}
The general aim of this paper is to supply a method to decide whether a
discrete system decoheres or not, and under what conditions decoherence
occurs, with no need of appealing to computer simulations to obtain the time
evolution of the reduced state. In particular, a lemma is presented as the
core of the method.

\end{abstract}
\maketitle

\section{Introduction}

At present, the study of quantum decoherence has acquired a central position
in the theoretical research on quantum mechanics. Although the orthodox
environment induced decoherence (EID) approach addresses decoherence in open
systems (\cite{Zeh-1970}-\cite{Zurek-2003}), many authors have stressed that
closed systems may also experience the phenomenon of decoherence
(\cite{Diosi-1}-\cite{Sicardi}; we have worked from this perspective in
\cite{SID-1}-\cite{CO-CSF}).

In order to show that the two approaches must not be conceived as rival or
alternative, but rather as complementary, we have developed a \textit{general
theoretical framework }for decoherence (\cite{CQG-General}, \cite{JPA-Sebi},
\cite{PLA}), which encompasses decoherence in open and closed systems.
According to this general framework, decoherence is just a particular case of
the general phenomenon of irreversibility in quantum mechanics
(\cite{Omnes-2001}, \cite{Omnes-2002}). In fact, since the quantum state
$\rho(t)$ follows a unitary evolution, it cannot reach a final equilibrium
state for $t\rightarrow\infty$. Therefore, if we want to explain the emergence
of non-unitary irreversible evolutions, a further element has to be added: we
must split the whole space $\mathcal{O}$ of all possible observables into a
relevant subspace $\mathcal{O}_{R}\subset\mathcal{O}$ and an irrelevant
subspace. Once the essential role played by the selection of the relevant
observables is clearly understood, the phenomenon of decoherence can be
explained in three general steps:

\begin{enumerate}
\item \textbf{First step:} The space $\mathcal{O}_{R}$ of relevant observables
is defined .

\item \textbf{Second step:} The expectation value $\langle O_{R}\rangle
_{\rho(t)}$, for any $O_{R}\in\mathcal{O}_{R}$, is obtained. This step can be
formulated in two different but equivalent ways:

\begin{itemize}
\item $\langle O_{R}\rangle_{\rho(t)}$ is computed as the expectation value of
$O_{R}$ in the unitarily evolving state $\rho(t)$.

\item A coarse-grained state $\rho_{R}(t)$ is defined by
\begin{equation}
\langle O_{R}\rangle_{\rho(t)}=\langle O_{R}\rangle_{\rho_{R}(t)}\qquad\forall
O_{R}\in\mathcal{O}_{R} \label{INT-00}%
\end{equation}
and its non-unitary evolution (governed by a master equation) is solved.
\end{itemize}

\item \textbf{Third step:} It is proved that $\langle O_{R}\rangle_{\rho
(t)}=\langle O_{R}\rangle_{\rho_{R}(t)}$ reaches a final equilibrium value:
\begin{equation}
\lim_{t\rightarrow\infty}\langle O_{R}\rangle_{\rho(t)}=\lim_{t\rightarrow
\infty}\langle O_{R}\rangle_{\rho_{R}(t)}=\langle O_{R}\rangle_{\rho_{\ast}%
}=\langle O_{R}\rangle_{\rho_{R\ast}}\text{\ \ \ \ \ \ \ \ \ }\forall O_{R}%
\in\mathcal{O}_{R} \label{INT-01}%
\end{equation}

\end{enumerate}

The final equilibrium state $\rho_{\ast}$ is obviously diagonal in its own
eigenbasis, which turns out to be the final equilibrium decoherence basis.
But, from eq. (\ref{INT-01}) it cannot be concluded that $\lim_{t\rightarrow
\infty}\rho(t)=\rho_{\ast}$; the mathematicians say that the unitarily
evolving quantum state $\rho(t)$ of the whole system\textit{ }has only a
\textit{weak limit},\textit{ }symbolized as:
\begin{equation}
W-\lim_{t\rightarrow\infty}\rho(t)=\rho_{\ast} \label{INT-03}%
\end{equation}
and equivalent to eq. (\ref{INT-01}). Physically this weak limit means that,
although the off-diagonal terms of $\rho(t)$ never vanish through the unitary
evolution, the system decoheres \textit{from an observational point of view},
that is, from the viewpoint given by any relevant observable $O_{R}%
\in\mathcal{O}_{R}$. From this general perspective, the phenomenon of
decoherence is relative because the off-diagonal terms of $\rho(t)$ vanish
only from the viewpoint of the relevant observables\textbf{ }$O_{R}%
\in\mathcal{O}_{R}$.

This general framework strictly applies when the limit of eq. (\ref{INT-01})
exists. This happen when the off-diagonal terms of the density matrix vanish
by destructive interference according to the Riemann-Lebesgue theorem. In
fact, when the relevant observables $O_{R}\in\mathcal{O}_{R}$ read%
\begin{equation}
O_{R}=\int_{0}^{\infty}O(\omega)|\omega)\,d\omega+\int_{0}^{\infty}\int
_{0}^{\infty}O(\omega,\omega^{\prime})|\omega,\omega^{\prime})\,d\omega
d\omega^{\prime} \label{INT-04}%
\end{equation}
where $\left\{  |\omega\rangle\right\}  $ is the eigenbasis of the
Hamiltonian, $|\omega)=|\omega\rangle\langle\omega|$, $|\omega,\omega^{\prime
})=\,|\omega\rangle\langle\omega^{\prime}|$, and $\,\left\{  |\omega
),|\omega,\omega^{\prime})\right\}  $ is a basis of $\mathcal{O}_{R}$, and the
states are linear functionals belonging to $\mathcal{O}_{R}^{\prime}$, the
dual of $\mathcal{O}_{R}$,
\begin{equation}
\rho=\int_{0}^{\infty}\rho(\omega)\left(  |\omega\rangle\langle\omega|\right)
^{\prime}\,d\omega+\int_{0}^{\infty}\int_{0}^{\infty}\rho(\omega
,\omega^{\prime})(\omega,\omega^{\prime}|\,d\omega d\omega^{\prime}
\label{INT-05}%
\end{equation}
where $\left\{  (\omega|,(\omega,\omega^{\prime}|\,\right\}  $ is the basis of
$\mathcal{O}_{R}^{\prime}$, then the expectation value of any observable
$O_{R}\in\mathcal{O}_{R}$ in the state $\rho(t)$ can be computed as
\begin{equation}
\langle O_{R}\rangle_{\rho(t)}=\int_{0}^{\infty}\overline{\rho(\omega
)}O(\omega)\,d\omega+\int_{0}^{\infty}\int_{0}^{\infty}\overline{\rho
(\omega,\omega^{\prime})}O(\omega,\omega^{\prime})\,e^{i\frac{\omega
-\omega^{\prime}}{\hbar}t}\,d\omega d\omega^{\prime} \label{INT-06}%
\end{equation}
When the function $\overline{\rho(\omega,\omega^{\prime})}O(\omega
,\omega^{\prime})$ is regular (precisely, when it is $\mathbb{L}_{1}$ in
variable $\nu=\omega-\omega^{\prime}$), the Riemann-Lebesgue theorem can be
applied to eq. (\ref{INT-06}). As a consequence, the second term vanishes and
$\langle O_{R}\rangle_{\rho(t)}$ converges to a stable value
\begin{equation}
\langle O_{R}\rangle_{\rho(t)}\longrightarrow\int_{0}^{\infty}\overline
{\rho(\omega)}O(\omega)\,d\omega=\langle O_{R}\rangle_{\rho_{\ast}}
\label{3-6}%
\end{equation}
where $\rho_{\ast}$ is diagonal in the eigenbasis of the Hamiltonian.

It is clear that the Riemann-Lebesgue theorem strictly applies only in cases
of continuous energy spectrum. However, we also know that we can use the
results coming from the continuous realm in quasi-continuous cases, that is,
in discrete models where (i) the energy spectrum is quasi-continuous, i.e.,
has a small discrete energy spacing, and (ii) the functions of energy used in
the formalism are such that the sums in which they are involved can be
approximated by Riemann integrals. These conditions are rather weak: in fact,
the overwhelming majority of the physical models studied in the literature on
dynamics, thermodynamics, quantum mechanics and quantum field theory are quasi-continuous.

The general aim of the present work is to supply a rigorous formulation of
this intuitive idea. In particular, we will develop a discrete analogue of the
Riemann-Lebesgue theorem, and this task will lead us to introduce a lemma in
terms of which it is possible to predict whether a discrete system decoheres
or not. For this purpose, the paper is organized as follows. In Section 2 we
will consider the three cases that can be distinguished regarding the discrete
analogue of the Riemann integral involved in the Riemann-Lebesgue theorem.
Section 3 will be devoted to formulate the conditions for the validity of the
discrete analogue of the Riemann integral. In Section 4, the lemma that
constitutes the core of the proposed method is presented, and its relationship
with the Discrete Fourier Transform is pointed out. Finally, in the
Conclusions we will stress the fact that, whereas the usual strategy in this
field is to rely on that computer simulations, our lemma makes possible to
draw conclusions without that strategy; moreover, our results are particularly
adequate to take advantage of the many mathematical methods of software
engineering based on the Discrete Fourier Transform.

\section{Three cases in the discrete analogue}

The Riemann-Lebesgue theorem $-$the mathematical expression of destructive
interference$-$ establishes that%
\begin{equation}
\text{ }f(\nu)\in\mathbb{L}_{1}\Longrightarrow\lim_{t\rightarrow\infty}%
\int_{0}^{1}d\nu f(\nu)e^{i\nu t}=0 \label{DRLT-01}%
\end{equation}
\ In order to obtain a version that can be used in the discrete case, let us
analyze the discrete analogue of the Riemann integral $R(t)$:%
\begin{equation}
R(t)=\int_{0}^{1}d\nu f(\nu)e^{i\nu t}\longrightarrow R_{D}(t)=\sum_{j=0}%
^{N}\frac{1}{N}f\left(  \frac{j}{N}\right)  e^{i\frac{j}{N}t} \label{DRLT-02}%
\end{equation}
where $0\leq j/N\leq1$,\ and $t$ is a dimensionless time. Since the function
$R_{D}(t)$ is a finite sum of sine functions $f\left(  \frac{j}{N}\right)
e^{i\frac{j}{N}t}$, it has a recurrence or Poincar\'{e} time. Given an initial
state $R_{D}(0)$, the Poincar\'{e} time $t_{P}$ is defined by $R_{D}%
(0)=R_{D}(t_{P})$. This means that%
\begin{equation}
\sum_{j=0}^{N}\frac{1}{N}f\left(  \frac{j}{N}\right)  \left(  e^{i\frac{j}%
{N}t_{P}}-1\right)  =0\quad\Longrightarrow\quad t_{P}=2\pi\label{DRLT-03}%
\end{equation}

Since the system comes back to the initial state when $t=t_{P}$ , there is no
rigorous discrete analogue of the Riemann-Lebesgue theorem. Nevertheless,
three possible situations can be distinguished:

\begin{enumerate}
\item If $N\rightarrow\infty$, then $\left\vert \frac{j+1}{N}-\frac{j}%
{N}\right\vert $ becomes infinitesimal and $t_{P}\rightarrow\infty$.
Therefore, this situation can be considered a continuous-spectrum case where
the Riemann-Lebesgue theorem can be applied.

\item If $N$\ is large, then $\left\vert \frac{j+1}{N}-\frac{j}{N}\right\vert
$ is very small. Then, the sum turns out to be close to the Riemann integral,
and the the situation can be approximated to the continuous-spectrum case
where the Riemann-Lebesgue theorem can be applied. This condition is satisfied
in a concrete example in \cite{Alimanas}: in spite of the fact that, strictly
speaking, a system with discrete spectrum never reaches equilibrium due to
Poincar\'{e} recurrence, that paper shows that, for times $t\ll t_{P}$, the
discrete spectrum can be approximated by a continuous spectrum where the
considered functions satisfy the usual conditions of regularity and integrability.

\item If $N$\ is not large, then $\left\vert \frac{j+1}{N}-\frac{j}%
{N}\right\vert $ is far from being infinitesimal, and the sum cannot be
approximated by a Riemann integral. Consequently, the Riemann-Lebesgue theorem
is not applicable because there is no destructive interference.
\end{enumerate}

Let us notice that the difference between cases 2 and 3 is not absolute, to
the extent that the precise criterion to decide when $N$ is large has not been
defined. In the following subsections such a criterion will be established.

\section{Conditions for the validity of the discrete analogue}

The problem is to find the conditions for the time $t_{F}$ such that
$R_{D}(t)\rightarrow0$ when $t\rightarrow t_{F}$: therefore, in the time-scale
$\left[  0,t_{F}\right]  $ it can be considered that $N$ is large enough to
make the continuous-spectrum approximation applicable. In order to face this
problem, we begin by consider a fixed time $t$ and by decomposing the
exponential of eq. (\ref{DRLT-02}) as
\begin{equation}
R_{D}(t)=\sum_{i=0}^{N}\frac{1}{N}f\left(  x_{i}\right)  e^{ix_{i}t}%
=\sum_{i=0}^{N}\frac{1}{N}f\left(  x_{i}\right)  \cos\left(  x_{i}t\right)
+i\sum_{i=0}^{N}\frac{1}{N}f\left(  x_{i}\right)  \sin\left(  x_{i}t\right)
\label{DRLT-04}%
\end{equation}
where the points $x_{i}=i/N$ belong to a discrete set $\left\{  x_{i}\right\}
$ with $i=0$ to $N$. Let us analyze the particular case where $f\left(
x_{i}\right)  =1$, precisely,%
\begin{equation}
R_{D}^{(1)}(t)=\sum_{i=0}^{N}\frac{1}{N}\cos\left(  x_{i}t\right)
+i\sum_{i=0}^{N}\frac{1}{N}\sin\left(  x_{i}t\right)  \label{DRLT-05}%
\end{equation}
In particular, we will consider the sums%
\begin{equation}
\frac{1}{N}\sum_{i=0}^{N}\cos\left(  x_{i}t\right)  =R_{D}^{(1C)}%
(t)\qquad\qquad\frac{1}{N}\sum_{i=0}^{N}\sin\left(  x_{i}t\right)
=R_{D}^{(1S)}(t) \label{DRLT-06}%
\end{equation}
We begin by considering $R_{D}^{(1C)}(t)$, because the case of $R_{D}%
^{(1S)}(t)$ will be analogous.

The sum $R_{D}^{(1C)}(t)$ vanishes when its terms cancel by pairs, that is,
when for any $x_{i}\in\left\{  x_{i}\right\}  $%
\begin{equation}
\cos\left(  x_{i}t\right)  +\cos\left(  x_{i}t+\pi\right)  =0 \label{DRLT-08}%
\end{equation}
where $x_{k}=x_{i}+\pi/t$ $\in\left\{  x_{i}\right\}  $. Since $x_{i}=i/N$,
this happens when
\begin{equation}
\frac{k}{N}=\frac{i}{N}+\frac{\pi}{t}\Rightarrow k-i=\frac{\pi}{t}N
\label{DRLT-09}%
\end{equation}
where $k-i$ $\in\mathbb{N}$ and $N\in\mathbb{N}$. However, since in general
$\pi/t\notin\mathbb{N}$, the condition of eq. (\ref{DRLT-09}) is not always
satisfied. Then, instead of requiring that the terms of $R_{D}^{(1C)}(t)$
cancel with each other exactly, we will only require that the corresponding
difference be small in the following sense:%
\begin{equation}
\left\vert \cos\left(  x_{i}t\right)  +\cos\left(  x_{i}t+\pi+\delta
_{j}\right)  \right\vert <\varepsilon\ll1 \label{DRLT-10}%
\end{equation}
where now $x_{k}=x_{i}+\pi/t+\delta_{i}/t\in\left\{  x_{i}\right\}  $ and%
\begin{equation}
\delta_{j}=\min_{i}\left\{  \delta_{i}\right\}  \qquad\text{with}\quad
\delta_{i}=x_{k}t-x_{i}t-\pi\quad\text{ }\left(  i=0...N\right)
\label{DRLT-11}%
\end{equation}
If $\delta_{j}<1$, the Taylor development of $\cos\left(  x_{i}t+\pi
+\delta_{j}\right)  $ leads to%
\begin{equation}
\left\vert \cos\left(  x_{i}t\right)  +\cos\left(  x_{i}t+\pi+\delta
_{j}\right)  \right\vert \simeq\left\vert \sin\left(  x_{i}t\right)
\delta_{j}+\cos\left(  x_{i}t\right)  \delta_{j}^{2}\right\vert <\varepsilon
\label{DRLT-14}%
\end{equation}
But, on the other hand,
\begin{align}
\left\vert \sin\left(  x_{i}t\right)  \delta_{j}+\cos\left(  x_{i}t\right)
\delta_{j}^{2}\right\vert  &  =\left\vert \delta_{j}\right\vert \left\vert
\sin\left(  x_{i}t\right)  +\cos\left(  x_{i}t\right)  \delta_{j}\right\vert
\nonumber\\
&  \leq\left\vert \delta_{j}\right\vert \left(  \left\vert \sin\left(
x_{i}t\right)  \right\vert +\left\vert \cos\left(  x_{i}t\right)  \right\vert
\left\vert \delta_{j}\right\vert \right)  \leq\left\vert \delta_{j}\right\vert
\left(  1+\left\vert \delta_{j}\right\vert \right)  \leq\left\vert \delta
_{j}\right\vert \label{DRLT-15}%
\end{align}
Then, if $\left\vert \delta_{j}\right\vert <\varepsilon\ll1$, from eqs.
(\ref{DRLT-14}) and (\ref{DRLT-15}) we obtain the condition of eq.
(\ref{DRLT-10}). Therefore, the condition of approximate cancelling is (see
eq. (\ref{DRLT-11}))
\begin{equation}
\left\vert \delta_{j}\right\vert <\varepsilon\ll1\qquad\text{with \ }%
\delta_{j}=x_{k}t-x_{j}t-\pi\label{DRLT-17}%
\end{equation}

Now we will express the condition of approximate cancelling of eq.
(\ref{DRLT-17}) in terms of the time $t$. Let us begin by noticing that the
condition is not satisfied for $t=0$, since $t=0\Rightarrow\delta_{j}%
=-\pi\Rightarrow\left\vert \delta_{j}\right\vert >\varepsilon$. Then, the
first condition is $t>0$. Now, by recalling that $x_{i}=i/N$, from the
expression of $\delta_{j}$ in eq. (\ref{DRLT-17}) we obtain%
\begin{equation}
k-j=\frac{\pi N}{t}+\frac{\delta_{j}N}{t} \label{DRLT-18}%
\end{equation}
But since $j,k\in\mathbb{N}$, and $j,k\in\left[  0,N\right]  $, then for
$j<k$
\begin{equation}
1\leq k-j\leq N\quad\Longrightarrow\quad1\leq\frac{\pi N}{t}+\frac{\delta
_{j}N}{t}\leq N \label{DRLT-19}%
\end{equation}
Then, if $\left\vert \delta_{j}\right\vert <\varepsilon\ll1$, eq.
(\ref{DRLT-19}) implies that
\begin{equation}
1\leq\frac{\pi N}{t}\leq N \label{DRLT-21}%
\end{equation}
Therefore, the condition $\left\vert \delta_{j}\right\vert <\varepsilon\ll1$
of approximate cancelling turns out to be%
\begin{equation}
\pi\leq t\leq\pi N \label{DRLT-22}%
\end{equation}

Up to this point we have proved that, for $\pi\leq t\leq\pi N$, $\cos\left(
x_{j}t\right)  $ approximately cancels with $\cos\left(  x_{j}t+\pi+\delta
_{j}\right)  $. Figure \ref{Fig-01} shows an example of this situation, where
point 1 is cancelled by point 9, point 2 by point 10, ... , point 8 by point
16. However, this is not the most general case, since the points cancel by
pairs only when $t=2\pi n$. In the general case there are points with no
counterpart to be cancelled. An example of this situation is shown in Figure
\ref{Fig-02}, where points $13$, $14$, $15$ and $16$ are not cancelled.%

\begin{figure}[t]
 \centerline{\scalebox{0.7}{\includegraphics{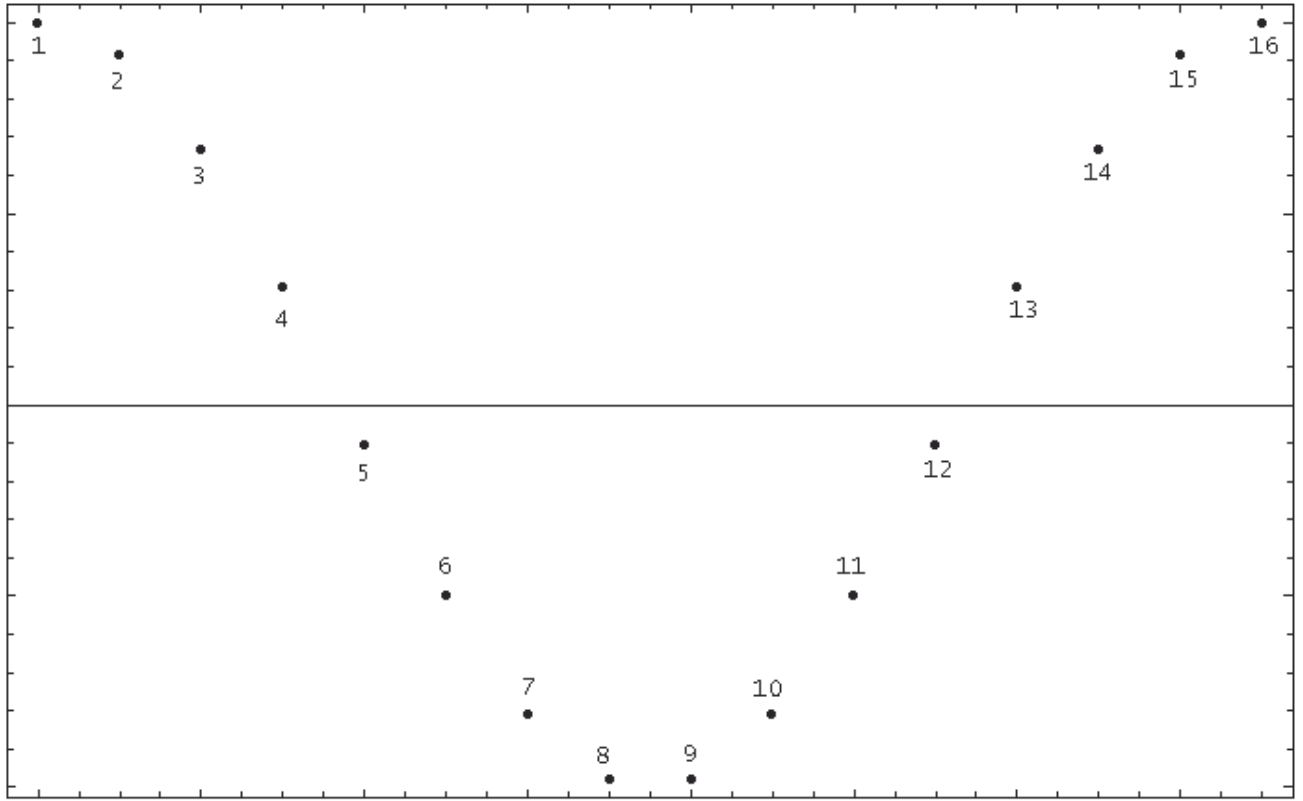}}}
\caption{Point 1 is cancelled by point 9, point 2 by point 10, ... , point 8
by point 16.}
 \label{fig 1}\vspace*{0.cm}
\end{figure}

\begin{figure}[t]
 \centerline{\scalebox{0.7}{\includegraphics{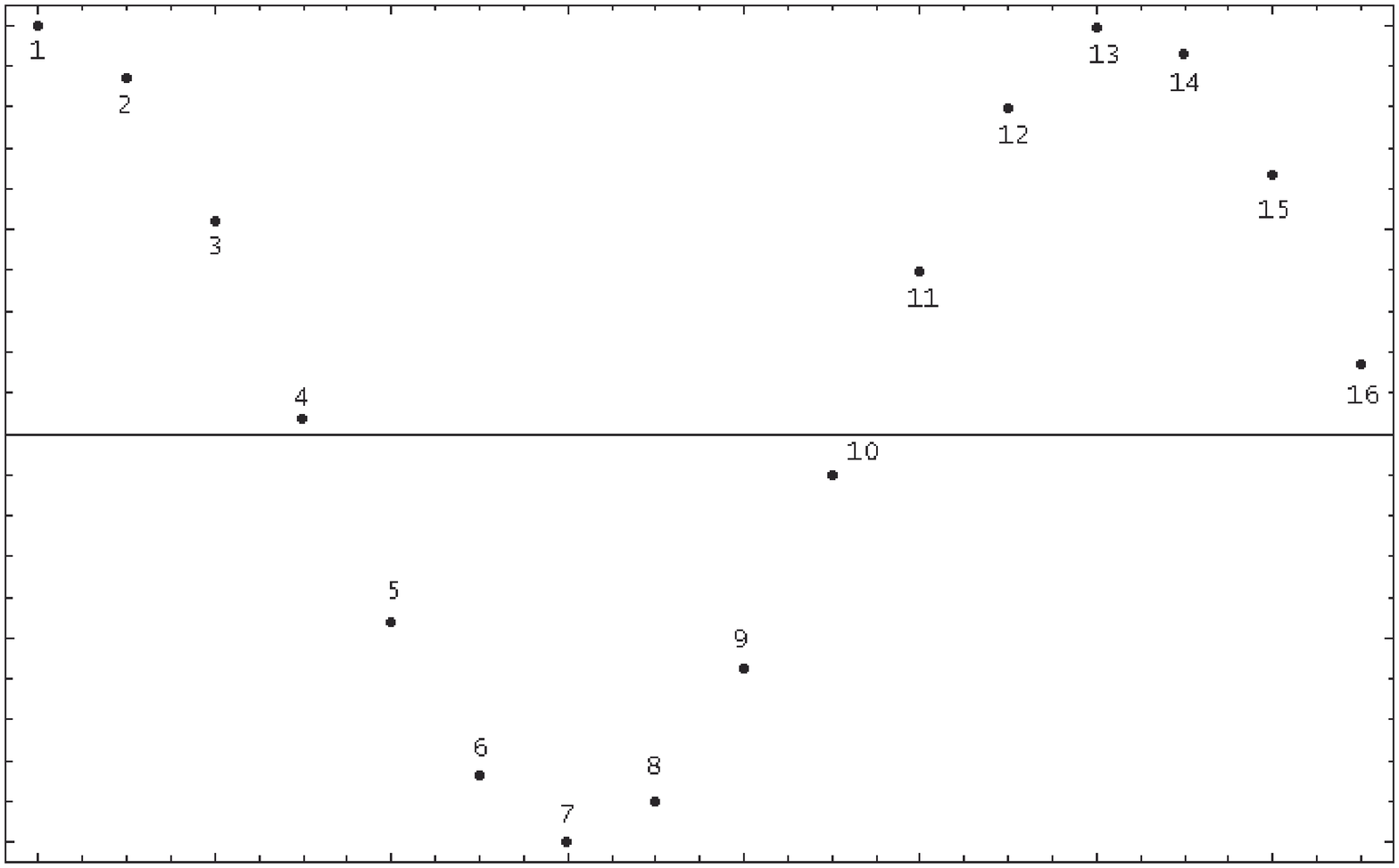}}}
\caption{Points $13$, $14$, $15$ and $16$ are not cancelled.}
 \label{fig 2}\vspace*{0.cm}
\end{figure}

In order to analyze this general situation, let us consider the
\textquotedblleft worst\textquotedblright\ case, when the points of a whole
half-period are not cancelled, namely, when $t=\left(  2n+1\right)  \pi$.
Since in $t$ there are $N+1$ points, in such a half-period there are $\left(
N+1\right)  /(2n+1)$ points, whose contribution $r_{\pi}(t)$ to the sum
$R_{D}^{(1C)}(t)$ is%
\begin{equation}
r_{\pi}(t)=\frac{1}{N}\sum_{i=0}^{\frac{N+1}{2n+1}-1}\cos\left(
x_{i}t\right)  \label{DRLT-23}%
\end{equation}
The upper boundary of this contribution is%
\begin{equation}
r_{\pi}(t)=\frac{1}{N}\sum_{i=0}^{\frac{N+1}{2n+1}-1}\cos\left(
x_{i}t\right)  <\frac{1}{N}\frac{N+1}{2n+1}\cong\frac{1}{2n+1}=\frac{\pi}{t}
\label{DRLT-24}%
\end{equation}
Then, the contribution $r_{\pi}(t)$ of the non cancelled points is irrelevant
when $r_{\pi}(t)=\pi/t<\varepsilon\ll1$, and this adds the condition%
\begin{equation}
t\gg\pi\label{DRLT-25}%
\end{equation}

Summing up, if we combine eqs. (\ref{DRLT-22}) and (\ref{DRLT-25}), we obtain
the condition on the time-scale that guarantees the approximate cancelling of
the terms of $R_{D}^{(1C)}(t)$:%
\begin{equation}
\text{If }\pi\ll t\leq\pi N\quad\Rightarrow\quad R_{D}^{(1C)}(t)<\varepsilon
\ll1 \label{DRLT-26-1}%
\end{equation}
By means of the same argument applied to $R_{D}^{(1S)}(t)$ we obtain an
analogous result that, when combined with eq. (\ref{DRLT-26-1}), leads to%
\begin{equation}
\text{If }\pi\ll t\leq\pi N\quad\Rightarrow\quad R_{D}^{(1)}(t)<\varepsilon
\ll1 \label{DRLT-26-2}%
\end{equation}

\section{A lemma for the application of the discrete analogue}

Up to this point we have studied the case $f\left(  x_{i}\right)  =1$; now we
will consider the case $f\left(  x_{i}\right)  \neq1$. In order to compute
$R_{D}(t)$ as defined by eq. (\ref{DRLT-04}), we have to ask a certain degree
of regularity to function $f\left(  x_{i}\right)  $. The first step consists
in splitting the set $\left\{  x_{i}\right\}  $ in $G$ subsets of $\left(
P+1\right)  $ consecutive points:%
\begin{equation}
\left\{  x_{i}\right\}  =\bigcup_{k=1}^{G}\left\{  x_{\left(  k-1\right)
\left(  P+1\right)  +1},...,x_{k\left(  P+1\right)  }\right\}  =\bigcup
_{k=1}^{G}X_{k} \label{DRLT-27}%
\end{equation}
Now we relabel the points $x_{j}\in X_{k}$: since $j=\left(  k-1\right)
\left(  P+1\right)  +1$ to $k\left(  P+1\right)  $, we can replace the index
$j$ by the index $r_{k}=j+\left(  1-k\right)  \left(  P+1\right)  -1$, and we
obtain
\begin{equation}
x_{j}\in X_{k}\longrightarrow x_{r_{k}}\in X_{k}\quad\text{ with }%
r_{k}=0,...,P \label{DRLT-28}%
\end{equation}
Then, we define

\begin{definition}
\textbf{.} \textit{Let} $\left\{  x_{i}\right\}  $ \textit{be a set of points
uniformly distributed (or equidistant), with} $i\in\left[  0,N\right]  $
\textit{and} $N\gg1$. The set $\left\{  x_{i}\right\}  $ \textit{is}
\textit{said to be quasi-continuous of class 1 if} $\exists G\in
\mathbb{N},\exists P\in\mathbb{N}$ such that $P\gg1$ and $\left\{
x_{i}\right\}  =\bigcup_{k=1}^{G}\left\{  x_{\left(  k-1\right)  \left(
P+1\right)  +1},...,x_{k\left(  P+1\right)  }\right\}  =\bigcup_{k=1}^{G}%
X_{k}$. \textit{The set }$X_{k}$\ \textit{is called the} $k$ \textit{component
of the quasi-continuous decomposition.}
\end{definition}

If the function $f(x_{r_{k}})$ is almost constant in $X_{k}$, i.e.%
\begin{equation}
f(x_{r_{k}})\cong C_{k}\text{ } \label{DRLT-29}%
\end{equation}
Then, we can define

\begin{definition}
\textbf{.} \textit{Let} $f(x_{i}):\mathbb{R}\rightarrow\mathbb{R}$ \textit{be
a discrete function defined over the quasi-continuous set} $\left\{
x_{i}\right\}  $ \textit{of class 1.} \textit{If for every component }$X_{k}$
\textit{of a quasi-continuous decomposition} $f(x_{r_{k}})\cong C_{k}$,
\textit{with} $x_{r_{k}}\in X_{k}$, we say that \ $f(x_{i})\in\mathcal{L}_{1}$\ .
\end{definition}

Therefore, when $f(x_{i})\in\mathcal{L}_{1}$, the discrete function $R_{D}(t)$
can be written as\
\begin{equation}
R_{D}(t)=\sum_{i=0}^{N}\frac{1}{N}f\left(  x_{i}\right)  e^{ix_{i}t}%
=\sum_{k=1}^{G}\frac{P}{N}\left(  \sum_{r_{k}=0}^{P}\frac{1}{P}f\left(
x_{r_{k}}\right)  e^{ix_{r_{k}}t}\right)  =\sum_{k=1}^{G}\frac{P}{N}%
C_{k}\left(  \sum_{r_{k}=0}^{P}\frac{1}{P}e^{ix_{r_{k}}t}\right)
\label{DRLT-30}%
\end{equation}
If we define the function%
\begin{equation}
R_{D}^{\left(  k\right)  }(t)=\sum_{r_{k}=0}^{P}\frac{1}{P}e^{ix_{r_{k}}t}
\label{DRLT-31}%
\end{equation}
then the discrete function $R_{D}(t)$ results
\begin{equation}
R_{D}(t)=\sum_{k=1}^{G}\frac{P}{N}C_{k}R_{D}^{\left(  k\right)  }(t)
\label{DRLT-32}%
\end{equation}
Under this form, the condition of eq. (\ref{DRLT-26-2}) obtained in the
previous subsection can be applied to each $R_{D}^{\left(  k\right)  }(t)$:%
\begin{equation}
\text{If }\pi\ll t\leq\pi P\quad\Rightarrow\quad R_{D}^{(k)}(t)<\varepsilon
\ll1 \label{DRLT-32 bis}%
\end{equation}
When this condition is satisfied, the sum $R_{D}(t)$ results
\begin{equation}
R_{D}(t)=\sum_{k=1}^{G}\frac{P}{N}C_{k}R_{D}^{\left(  k\right)  }%
(t)<\sum_{k=1}^{G}\frac{P}{N}C_{k}\varepsilon_{k}\leq\sum_{k=1}^{G}\frac{P}%
{N}C\varepsilon=\frac{PG}{N}C\varepsilon=C\varepsilon\label{DRLT-33}%
\end{equation}
where $\varepsilon=\max_{k}\left\{  \varepsilon_{k}\right\}  $ and $C=\max
_{k}\left\{  C_{k}\right\}  $. As a consequence, if we consider that
$t_{P}=2\pi$, we have proved that

\textbf{Lemma 1.} \textit{Let} $f(x_{i})$ \textit{be} \textit{defined over the
quasi-continuous set} $\left\{  x_{i}\right\}  $ \textit{of class 1, with}
$i=1...N$\textit{. If }$f(x_{i})\in\mathcal{L}_{1}$,\textit{ then}%
\begin{equation}
\ \lim_{t\longrightarrow t_{P}/2}\sum_{i=0}^{N}\frac{1}{N}f\left(
x_{i}\right)  e^{ix_{i}t}\cong0 \label{DRLT-34}%
\end{equation}
\medskip

There are different kinds of functions for which the sum $R_{D}(t)$ vanishes
and that could be characterized by further lemmas, but we will not consider
those cases now. Nevertheless, a practically useful remark is in order. The
condition of eq. (\ref{DRLT-29}) (that the function $f(x_{i})$ be
approximately constant in each element $X_{k}$ of the partition) can be
expressed under a matematically more elegant form. Given a function $f(x_{i}%
)$, its \textit{Discrete Fourier Transform} (DFT), as used in signal analysis
(\cite{Fourier-1}-\cite{Fourier-4}), is defined as
\begin{equation}
\tilde{f}(t)=\sum_{i=0}^{N}\frac{1}{N}f\left(  x_{i}\right)  e^{ix_{i}t}
\label{DRLT-37}%
\end{equation}
This result may be very useful in practice, in particular in cases in which
Lemma 1 is difficult to be applied. In fact, when we realize that the sum
$R_{D}(t)$ corresponding to the function $f\left(  x_{i}\right)  $ is
precisely the DFT of $f\left(  x_{i}\right)  $, we can use all the properties
of DFT $-$as linearity, symmetry, time-shifting, frequency-shifting, the
time-convolution theorem, and the frequency-convolution theorem$-$ to study
$R_{D}(t)$. Moreover, we can take advantage of the large amount of software
designed to compute DFT and profusely used in physics and engineering. All
these resources, which are standard tools in signal analysis, may prove to be
extremely useful for studying decoherence in discrete models.

\section{Conclusions}

In this paper we have offered a discrete analogue of the Riemann-Lebesgue
theorem and, on this basis, we have introduced a lemma relevant for discrete
models, which provides a criterion for deciding whether or not the system
decoheres with no need of numerical simulations. Moreover, we have shown how
the large amount of mathematical methods of software engineering based on the
Discrete Fourier Transform can be used to predict decoherence in discrete models.

\section{Acknowledgments}

This research was partially supported by grants of the University of Buenos
Aires, the CONICET and the FONCYT of Argentina.


\begin{thebibliography}{99}                                                                                               %


\bibitem {Zeh-1970}H. D. Zeh, \textit{Found. Phys.}, \textbf{1}, 69, 1970

\bibitem {Zeh-1973}H. D. Zeh, \textit{Found. Phys.}, \textbf{3}, 109, 1973.

\bibitem {Zurek-1982}W. H. Zurek, \textit{Phys. Rev. D}, \textbf{26}, 1862, 1982.

\bibitem {Zurek-1993}W. H. Zurek, \textit{Progr. Theor. Phys}., \textbf{89},
281, 1993.

\bibitem {Paz-Zurek}J. P. Paz and W. Z., \textquotedblleft Environment-induced
decoherence and the transition from quantum to classical\textquotedblright, in
Dieter Heiss (ed.), \textit{Lecture Notes in Physics, Vol. 587}, Springer,
Heidelberg-Berlin, 2002.

\bibitem {Zurek-2003}W. H. Zurek, \textit{Rev. Mod. Phys}., \textbf{75}, 715, 2003.

\bibitem {Diosi-1}L. Diosi, \textit{Phys. Lett. A}, \textbf{120, }377, 1987.

\bibitem {Diosi-2}L. Diosi, \textit{Phys. Rev. A}, \textbf{40, }1165, 1989.

\bibitem {Milburn}G. J. Milburn, \textit{Phys. Rev. A}, \textbf{44, }5401, 1991.

\bibitem {Penrose}R. Penrose, \textit{Shadows of the Mind},\textit{ }Oxford
University Press, Oxford, 1995.

\bibitem {Casati-Chirikov-1}G. Casati and B. Chirikov, \textit{Phys. Rev.
Lett.}, \textbf{75,} 349, 1995.

\bibitem {Casati-Chirikov-2}G. Casati and B. Chirikov, \textit{Physica D},
\textbf{86, }220, 1995.

\bibitem {Adler}S. Adler, \textit{Quantum Theory as an Emergent Phenomenon}%
,\textit{ }Cambridge University Press, Cambridge, 2004.

\bibitem {Bonifacio}R. Bonifacio, S. Olivares, P. Tombesi and D. Vitali,
\textit{Phys. Rev. A}, \textbf{61}, 053802, 2000.

\bibitem {Ford}G. W. Ford and R. F. O'Connell, \textit{Phys. Lett. A},
\textbf{286}, 87, 2001.

\bibitem {Frasca}M. Frasca, \textit{Phys. Lett. A}, \textbf{308}, 135, 2003.

\bibitem {Sicardi}A. C. Sicardi Shifino, G. Abal, R. Siri, A. Romanelli and R.
Donangelo, \textquotedblleft Intrinsic decoherence and irreversibility in a
quasiperiodic kicked rotor\textquotedblright, arXiv:quant-ph/0308162, 2003.

\bibitem {SID-1}M. Castagnino, \textit{Int. J. Theor. Phys.}, \textbf{38},
1333, 1999.

\bibitem {CL-PRA-2000}M. Castagnino and R. Laura, \textit{Phys. Rev. A},
\textbf{62}, 022107, 2000.

\bibitem {CL-IJTP-2000}M. Castagnino and R. Laura, \textit{Int. Jour. Theor.
Phys.}, \textbf{39}, 1767, 2000.

\bibitem {CO-IJTP}M. Castagnino and O. Lombardi,\textit{ } \textit{Int. Jour.
Theor. Phys.}, \textbf{42}, 1281, 2003.

\bibitem {Cast-2004}M. Castagnino, \textit{Physica A}, \textbf{335}, 511, 2004.

\bibitem {COrd}M. Castagnino and A. Ordo\~{n}ez, \textit{Int. Jour. Theor.
Phys.}, \textbf{43}, 695, 2004.

\bibitem {Cast-2005}M. Castagnino, \textit{Braz. Jour. Phys.}, \textbf{35},
375, 2005.

\bibitem {CO-PR(DT)}M. Castagnino and O. Lombardi,\textit{ Phys. Rev. A},
\textbf{72}, 012102, 2005.

\bibitem {CO-PS}M. Castagnino and O. Lombardi,\textit{ Phil. Scie}.,
\textbf{72}, 764, 2005.

\bibitem {Cast-2006}M. Castagnino, \textit{Phys. Lett. A}, \textbf{357}, 97, 2006.

\bibitem {CG-CL}M. Castagnino and M. Gadella, \textit{Found Phys}.,
\textbf{36}, 920, 2006.

\bibitem {CO-CSF}M. Castagnino and O. Lombardi,\textit{ Chaos, Sol. Frac}.,
\textbf{28}, 879, 2006.

\bibitem {CQG-General}M. Castagnino, S. Fortin, R. Laura and O. Lombardi,
\textit{Class. Quantum Grav.}, \textbf{25}, 154002, 2008.

\bibitem {JPA-Sebi}M. Castagnino, S. Fortin and O. Lombardi, \textit{Jour.
Phys. A: Math. and Theo}r., \textbf{43}, 065304, 2010.

\bibitem {PLA}M. Castagnino, S. Fortin and O. Lombardi, \textit{Mod. Physics
Lett.} A, \textbf{25}, 1431, 2010.

\bibitem {Omnes-2001}R. Omn\`{e}s, ``Decoherence: an irreversible process'',
arXiv:quant-ph/0106006, 2001.

\bibitem {Omnes-2002}R. Omn\`{e}s, \textit{Phys. Rev. A}, \textbf{65}, 052119, 2002.

\bibitem {Alimanas}F. Gaioli, E. Garc\'{\i}a-\'{A}lvarez and J. Guevara,
\textit{Int. J. Theor. Phys.}, \textbf{36}, 2167, 1997.

\bibitem {Fourier-1}S. Bochner and K. Chandrasekharan, \textit{Fourier
Transforms}, Princeton University Press, London, 1949.

\bibitem {Fourier-2}E. Oran Brigham, \textit{The Fast Fourier Transform and
its Applications}, Prentice-Hall, New Jersey, 1988.

\bibitem {Fourier-3}W. Press , J. Teukolsky , E. Vetterling and B. Flanery,
\textit{Numerical Recipes in Fortran}, Cambridge University Press, Cambridge, 1994.

\bibitem {Fourier-4}J. Kauppinen and J. Partanen, \textit{Fourier Transforms
in Spectroscopy}, Wiley-VCH, Berlin, 2001.
\end{thebibliography}
\end{document}